\newcommand{\la}{\langle}
\newcommand{\ra}{\rangle}
\newcommand{\lla}{\langle \langle}
\newcommand{\rra}{\rangle \rangle}
\renewcommand{\d}{\partial}
\newcommand{\beq}{\begin{eqnarray}}
\newcommand{\eeq}{\end{eqnarray}}
\newcommand{\sbeq}{\begin{subeqnarray}}
\newcommand{\seeq}{\end{subeqnarray}}

\newcommand{\btem}{\bibitem}

\newcommand{\bfq}{\mbox{{\boldmath $q$}}}

\newcommand{\THK}{T. Hatsuda\ and \ T. Kunihiro\ }
\newcommand{\PR}{Phys. Rev. }
\newcommand{\NPA}{Nucl. Phys. {\bf A}}

\documentclass[a4paper]{article}
\title{Chiral Transition and the Scalar and Vector Correlations}

\author{Teiji Kunihiro\\ 
Yukawa Institute for Theoretical Physics,
 Kyoto University, \\ 
Sakyo-ku, Kyoto 606-8502, Japan}

\begin{document}

\maketitle

\begin{abstract}
The properties of the scalar and vector correlations 
 in the hot and/or dense hadronic matter close to 
chiral transition are discussed. Presuming  that
the linear realization of chiral symmetry will become
appropriate at least near the critical point, 
we argue that the strength function in the
$I=J=0$ channel will soften near the critical 
point, and  the sigma meson will accordingly become a clearer
resonance in hot and/or dense medium than in the free space.
It is shown that the steep rise of the
baryon-number susceptibility $\chi_B$ around the critical
 point  seen in the lattice simulations 
may suggest
 that the interactions between quarks
in the vector channel become week near the critical 
point; this implies that the peculiar behavior of the 
$\chi_B$ does not necessarily imply a proliferation of
baryons and antibaryons (Skyrmions) which is though 
a natural scenario of chiral transition 
in the non-linear realization of chiral symmetry.  
\end{abstract}

\section{The sigma meson and the soft modes}

Chiral transition is a phase transition of the QCD vacuum with 
$\la \bar {q}q\ra$ being the order parameter.
When exploring a phase transition  in any physical 
system, the study of fluctuations of physical quantities, especially
ones related
to the order parameter is as important as that of the phase diagram 
for the system  in equilibrium. 
The fluctuations of observables are also related with 
dynamical phenomena such as  the transport properties of the system.

If a phase transition is of 2nd order or {\em weak} 1st order,
there exist soft modes, which are
the fluctuations of the order parameter as is well known 
in condensed matter physics\cite{soft} and nuclear physics
\cite{picon}.
 For the chiral transition,
the fluctuation of the order parameter 
$\langle(\bar{q}q)^2\rangle$ 
is a scalar-isoscalar meson which one calls the 
$\sigma$-meson.
The $\sigma$ meson may become
 the soft mode of chiral transition at $T\not=0$ and/or
$\rho_B\not=0$\cite{softnjl,physrep}.
In fact, the existence of the sigma meson as the quantum fluctuation
of the order parameter of the chiral transition is still 
controversial\cite{sigmayitp}, 
though the recent active studies on the
 phase-shift analysis of the $\pi$-$\pi$ scattering in the
$I=J=0$ channel, i.e., the sigma channel, have been confirming
 the existence of the pole deep in the second Riemann sheet
with the real part ranging from 400 MeV to 800 MeV
\cite{pipiyitp,pipi}. If this pole is identified with the
quantum fluctuation of the amplitude of the chiral
condensate, it may imply that 
the linear realization of chiral symmetry is 
appropriate in the resonance energy region, although
the chiral perturbation theory based on 
 the non-linear chiral Lagrangians\cite{chipert}
 works well in the rather low-energy region.

The first half of this report is concerned with 
 the sigma meson and the precursory soft modes for 
 chiral transition in nuclear medium, 
and is essentially a recapitulation
of the talk presented at Tohoku university, a week before of
the present workshop; the manuscript for the proceedings
is now on the net\cite{tohoku}.
Referring to the above proceedings for the detail of the
content, here I only give a brief summary of the report and
 some comments: 

(1) The existence of the $\sigma$ meson as 
{\em the quantum fluctuation of the
 order parameter} of the chiral transition
 accounts for various  phenomena in hadron physics
which otherwise remain mysterious\cite{elias,physrep,supple}.

(2) There have been accumulation of 
 experimental evidence of a low-mass pole in the $\sigma$ channel in
 the pi-pi scattering matrix\cite{pipiyitp}. 
It should be emphasized that for obtaining this result,
it is essential
 to respect chiral symmetry, analyticity and crossing symmetry
 even in an approximate way as in the $N/D$ method\cite{igi}.
 
(3)Partial restoration of chiral symmetry in hot and dense medium
 leads to an
enhancement in the spectral function
 in the $\sigma$ channel near the $2m_{\pi}$ threshold\cite{chiku}.
  Such an enhancement has been observed 
 in the reaction
\[
{\rm A}(\pi^{+}, (\pi^{+}\pi^{-})_{I=J=0}){\rm A}' 
\]
by CHAOS collaboration\cite{chaos}, which  might be 
 an experimental evidence of the partial restoration 
of chiral symmetry in heavy nuclei\cite{hks,chanf}:
The conventional  approach without incorporating a
partial restoration of chiral symmetry in the nuclear
medium\cite{convention} failed to reproduce the CHAOS data.

(4)  The  spectral enhancement near the 
$2m_{\pi}$ threshold in the
 $\sigma$ channel is predicted irrespective of the
linear and nonlinear realization of chiral symmetry
provided that the
 possible reduction of the quark condensate or $f_{\pi}$
 is taken into account\cite{jido}.

(5) One should confirm 
that the near $2m_{\pi}$-threshold enhancement observed in 
the ($\pi^{+}$, $\pi^{+}\pi^{-}$) reactions by CHAOS collaboration
 is surely due to a partial restoration of chiral symmetry in nuclear
 medium by other means\cite{hks}.
For this purpose,
the strength function in the $\sigma$ channel in the wider 
{\em two-dimensional}
 $(\omega, q)$ plane should be measured.
To obtain such strength functions in the  $(\omega, q)$ plane,
various nuclear and electro-magnetic
 probes as well as heavy-ion collisions should can be
utilized;
for instance, photo- or electro-production\cite{supple} 
of the $\sigma$ as well
 as the production 
by (d, $^{3}$He) and (d, t) reactions are interesting
\cite{hks,hiren}.

(6) Such  experiments with nuclear targets
for exploring the possible restoration of chiral
 symmetry in nuclear medium will
 automatically give  clearer confirmation of the existence of the
 $\sigma$ meson than is done in the free space.

(7) It should be  emphasized
 that in any hadronic medium where the baryon density is
finite, 
there arises a scalar-vector mixing\cite{weldon,supple,qnumk},
\beq
\sigma \leftrightarrow \gamma,\,  \omega,\nonumber
\eeq
 as is familiar in the $\sigma$-$\omega$ model\cite{walecka}, 
which may cause a possible
softening of the spectral functions at finite three momenta
 ${\bfq}$, even in the vector channel,
 due to the softening of
 that in the sigma channel associated with the partial
restoration of chiral symmetry in the hadronic medium.
Such a softening may reflect in the spectral functions 
extracted in any experiment to try to
see the spectral functions in the vector channel by seeing
 the lepton pairs from heavy-ion collisions such as 
CERES/NA45\cite{ceres}, 
proton-\cite{ozawa}, electro- or $gamma$-nucleus reactions
 and so on.

(8) The formation of $\sigma$ mesic nuclei
by (d, $^{3}$He) and (d, t) reactions are proposed as was done to
produce the deeply-bound pionic atoms\cite{hiren}.\,
 To identify the $\sigma$ meson and the spectral function in 
 that
 channel, detecting 2$\pi^{0}$ and lepton pairs with $q\not=0$ are
 interesting\cite{supple}. 

\section{Enhancement of the baryon-number susceptibility and 
the density fluctuations}

The discussions in the preceding section
 put an emphasis on the sigma meson as a
quantum fluctuation of the amplitude of the 
order parameter; this presumed that the 
 linear realization of the chiral symmetry is appropriate 
 at least near the critical point.
I personally take it for granted that the linear
realization is natural at least in the vicinity of the chiral
critical point.

Nevertheless some stick to the nonlinear realization even in the
vicinity of the critical point of the chiral transition.
If the non-linear realization is appropriate even in the
vicinity of the critical point,
the chiral restoration might be associated 
with an anomalous proliferation of the baryons and 
the anti-baryons as Skyrmions and anti-Skyrmions
near the critical point at finite 
temperature\cite{detar}, which may account for a steep rise 
of the baryon-number susceptibility,
$\chi _B$ \cite{qnum} obtained in the lattice 
simulations\cite{gott,lattice}.

It has also been suggested that the vector-realization 
\cite{harada,rho}, which 
 is based on the non-linear sigma model with 
 the vector mesons incorporated based on the ansatz of the 
 hidden-local symmetry\cite{bando} and
 similar to the Georgi's vector limit\cite{georgi},
could be realized; then the 
 decrease of the vector meson masses was conjectured\cite{rho}.

In the second half of the present report, I showed
 that the baryon-number susceptibility 
 give some information on the {\em vector}
correlations at finite temperature\cite{qnumk}, thereby
 the steep rise of $\chi _B$ seen in the lattice simulations
 may be accounted for without sticking to the non-linear realization
 of chiral symmetry.

The baryon-number susceptibility $\chi_B$  
is the measure of the response of the baryon
 number density 
\[
\rho_B=\sum _{\i=1\sim N_f}\rho_i\]
to infinitesimal changes in the quark chemical potentials $\mu _i$
\cite{qnum,qnumk}:
\beq
\chi _B(T,\mu)=
\bigl[\sum_{i=1}^{N_f}\frac {\partial} {\partial \mu_i}
\bigr]( \sum_{i=1}^{N_f}\rho_i)
=\lla {N}_B^2\rra/VT,
\eeq
where $ {N}_B $ is the baryon-number operator 
given by
$ N_B\equiv \sum_{i=1}^{N_f} N_i,$
with
 \beq
\rho _i={\rm Tr} N_i\exp [-\beta( H-\sum _{i=u,d}\mu _i
 N_i)]/V\equiv\lla N_i\rra/V
\eeq
 the $i$-th quark-number density,
 $V$ the volume of the system and $\beta=1/T$. 

It is readily recognized that $\chi_B$ is 
 the density-density correlation which is
 nothing but the 
 0-0 component of the vector-vector  correlations\cite{qnumk};
\beq
\chi _B(T,\mu_q)=\beta \int d{\bf x}S_{00}(0,{\bf x}),
\eeq
 where
\[
S_{\mu \nu}(t,{\bf x})=
\lla j_{\mu}(t,{\bf x})j_{\nu}(0,{\bf 0})\rra,\]
with
\[
j_{\mu}(t,{\bf x})= \bar {q}(t,{\bf x}) \gamma _{\mu}q(t,{\bf x})\]
being the  current operator.
Using the fluctuation-dissipation theorem, one has
\beq
\chi _B(T,\mu_q)=-\lim _{k\to 0}L(0,{\bf k}),
\eeq 
where $L(\omega, {\bf k})$ is the longitudinal component 
of the retarded Green's function or the response function in the
 vector channel;
\[
R _{\mu \nu}(\omega,k)={\rm F.T.}(-i\theta(t)\lla[j_{\mu}(t,{\bf x}),
\ j_{\nu}(0,{\bf 0})]_{-}\rra).\]
These formula clearly show the relevance of the
 vector correlations to the baryon-number susceptibility.

I also discussed the density 
fluctuations around the critical point of the chiral transition at
finite temperature $T$ and baryon density $\rho_B$.
We notice that the  baryon-number susceptibility 
at $\rho_B\not=0$  is 
 related with the (iso-thermal) compressibility of the system
\cite{qnumk},
\beq
\kappa _{_{T}}\equiv -N_B^{-1}(\d V/\d\mu )_{T,N_B}
=\frac {\chi_B}{\rho ^2},
\eeq
 which tells  
that if $\chi_B$ is large and so is the density fluctuation, 
the system is easy to  compress.
One can then see that the stronger interaction in the vector
 channel suppress $\chi _B$.

This part was largely based on a previous report by
myself\cite{conf00}, 
so I only give a brief summary of this part here, referring to
\cite{conf00} for the details:

(1) The baryon-number susceptibility $\chi_B$ as an
 observable which 
reflects the confinement-deconfinement and the chiral
 phase transitions in hot and/or dense hadronic matter\cite{qnum}.
   
(2) The suppression 
of $\chi_B$ at low temperatures and steep rise around the
critical temperature as shown in the lattice QCD may be roughly 
attributed to the confinement-deconfinement transition\cite{lattice}.
Nevertheless  such a behavior  of $\chi_B$ is also affected by the
chiral transition\cite{qnumk}.

(3) Noting that $\chi_B$ is a measure 
of the rate of the density fluctuation in the 
system,  one can see that 
the chiral transition at finite chemical potential 
especially leads to an interesting phenomenological consequence to 
$\chi_B$. When the vector coupling is small, the chiral transition 
at low temperatures is of first order 
in the density direction\cite{asakawa,instanton,random}.
This  implies a divergent behavior of $\chi_B$, 
accordingly a huge density fluctuations\cite{conf00}.

(4) A large enhancement of the
fluctuation can be also expected  
for the scalar density fluctuations
due to the scalar-vector mixing at finite density mentioned 
above\cite{qnumk,conf00}.
Such a large enhancement may leads to an enhancement of the
sigma-meson production due to the scalar-vector mixing 
at finite density.
The above phenomena all
 have relevance to experiments to be done in RHIC
and LHC\cite{kumagai,phenomena}. 

(5) The nature of the chiral transition as to the
first order or not etc 
is sensitively dependent on the strength of the
vector coupling\cite{asakawa}. 
An analysis of the  lattice data suggests that the vector 
coupling is small 
in comparison with the scalar 
coupling at high temperature\cite{qnumk,boyd}.
  
(6) The susceptibility $\chi_B$ is nothing but 
the generalized susceptibility 
$\chi(\omega,k)$ at $\omega = k =0$.
  One should examine $\chi(\omega,k)$
 in the whole region of $\omega $ and $k$ to get 
 more information about the vector correlations 
and the density fluctuations theoretically and experimentally. 

\vspace{.5cm}
\begin{Large}
\begin{center}
{\bf Acknowledgments}
\end{center}
\end{Large}

I thank the organizers of this
 symposium for inviting me to the symposium which 
 was organized to celebrate the retirement of
 Prof. I.-T. Cheon from Yonsei University.
I was very much pleased that I had a chance to 
present a talk at such a memorable symposium.
Some part of this  report is based on the works done 
in collaboration with T. Hatsuda, D. Jido and H. Shimizu,
to whom I am grateful.
This work is partially supported  by the Grants-in-Aid of
the Japanese Ministry of Education, Science and Culture
(No. 12640263 and 12640296).


\end{document}